# Anisotropic Transport for Parabolic, Non-Parabolic and Linear Bands of Different Dimensions

Shuang Tang*[1], Mildred S. Dresselhaus[2,3]

[1]Department of Materials Science and Engineering, Massachusetts Institute of Technology, Cambridge, MA, 02139-4037, USA; *tangs@mit.edu

[2]Department of Electrical Engineering and Computer Science, Massachusetts Institute of Technology Cambridge, MA, 02139-4037, USA;

[3]Department of Physics, Massachusetts Institute of Technology Cambridge, MA, 02139-4037, USA;

**Abstract:** Anisotropic thermoelectrics is a very interesting topic among recent research. The transport distribution function plays the central role on modeling the anisotropic thermoelectrics. The methodology of numerical integrations is used in previous literature on anisotropic transport, which does not capture the sharp change of transport distribution function and density of states at band edges. However, the sharp change of transport distribution function and density of states at band edges plays the central role in enhancing the thermoelectric performance. Thus, an analytical methodology that is robust on modeling the sharp change of transport distribution function and density of states at a band edges is needed. To our best knowledge, there has not been a paper giving the systematic study on the analytical models of anisotropic transport distribution function for different kinds of band valleys in different dimensions under different assumptions. Therefore, the main focus of this present paper is to develop such a robust analytical methodology on modeling the anisotropic transport distribution function. So our contributions are 1) we have developed a systematic method on model the anisotropic transport distribution function, for 3D, 2D and 1D systems, in parabolic, non-parabolic and linear dispersion relations, under both the relaxation time approximation and the mean free path approximation; 2) we have found that the Onsage's relation of transport can be violated under certain conditions; 3) we have compared our newly developed methodology with the traditional used numerical methodology.

Thermoelectrics (TE) has been recently intensively focused on by researchers. The performance of thermoelectricity generator and thermoelectric cooling can be characterized by the figure of merit $ZT=T\sigma S^2(\kappa_e+\kappa_L)^{-1}$, where $\sigma$, $S$, $\kappa_e$ and $\kappa_L$ are electrical conductivity, Seebeck coeffi-cient, electronic thermal conductivity and lattice thermal conductivity, respectively, which all reduce to scalars in isotropic materials. It can be seen that three out of the four transport quantities characterizing *ZT* are from electronic transport. As a matter of fact, most good TE materials are electronically anisotropic, such as $Bi_2Te_3$ and $Bi_{1-x}Sb_x$. Even isotropic TE/electronic materials, such as $Si_{1-x}Ge_x$, PbTe, PbS, also have anisotropic carrier-pockets. Thus, to calculate the electronic transport quantities associated anisotropic carrier-pockets in a simple physical way is very important for the research of optimizing TE/electronic performance. Good efforts have been made using numerical integrations for specific materials systems. However, the TE/electronic optimization problem requires closed-formed and physical expression of these transport quantities, such that the materials-searching and the conditions-optimization can be carried out among the huge number of material candidates and the various parameters, such as temperature *T*, Fermi level $E_f$, etc.

We recall that under the relaxation time approximation of Boltzmann equation,

$$\boldsymbol{\sigma} = e^2 \mathbf{I}_{[k=0]}, \tag{1}$$

$$\mathbf{S} = k_B e^{-1} \mathbf{I}_{[k=1]} \mathbf{I}_{[k=0]}^{-1}, \tag{2}$$

$$\boldsymbol{\kappa}_e = \boldsymbol{\kappa}_0 - T \mathbf{S}^T \boldsymbol{\sigma} \mathbf{S} \tag{3}$$

and

$$\boldsymbol{\kappa}_0 = T \cdot k_B^2 \mathbf{I}_{[k=2]}, \tag{4}$$

where

$$\mathbf{I}_{[k]} = \int (-\partial f_0 / \partial E) \boldsymbol{\Xi}(E)(E - E_f / k_B T)^k dE \tag{5}$$

and

$$\boldsymbol{\Xi}(E) = \sum_\mathbf{k} \delta(E - E_\mathbf{k}) \mathbf{v} \otimes \boldsymbol{\tau} \mathbf{v} \tag{6}$$

is the transport distribution tensor, and $\mathbf{v}$, $\boldsymbol{\tau}$ and $f_0$ are the carrier group velocity, the relaxation time tensor and the Fermi distribution, respectively. It is clear that the transport distribution tensor $\boldsymbol{\Xi}(E)$ plays the central role in determining all these electronic transport quantities. However, the complex calculations of inner- and outer-product between different ranks of tensors have always been misunderstood and improperly assumed in literatures that make numerical calculations of the transport distribution tensor for anisotropic systems.

Scheidenmatel et al.[1], Lee et al.[2], and Yavorsky et al.[3], have calculated the anisotropic transport distribution tensor with numerical integrations based on first-principle results for the specific material of intrinsic bulk $Bi_2Te_3$ without doping, by assuming that the relaxation time is a constant, which has given important references for this bulk material of intrinsic bulk $Bi_2Te_3$. However, these numerical methodologies take heavy computations, and is not easy for implement in the materials-searching and the conditions-optimization problems of TE/electronics. Teramoto et al.[4,5] have numerically calculated the anisotropic transport tensor in bulk Bi and bulk $Bi_{1-x}Sb_x$. However, a strong assumption is assumed that $\mathbf{v} \otimes \boldsymbol{\tau}\mathbf{v} = \boldsymbol{\tau}\mathbf{v} \otimes \mathbf{v}$, without further validation. Bies et al.[6] have made a remarkable progress on giving a relatively simpler expression for 3D and 2D parabolic bands with the dispersion in the form of $E(\mathbf{k}) = (h^2/2)\mathbf{k}^T \mathbf{M}^{-1} \mathbf{k}$, assuming that $\boldsymbol{\tau}(\mathbf{k}) = \tau_0[E(\mathbf{k})]\mathscr{U}$, where $\tau_0$ is a scalar function of $E$, and $\mathscr{U}$ is a dimensionless constant matrix. However, cases for the non-parabolic bands that happens at the $L$ points of Bi, $Bi_{1-x}Sb_x$, PbTe, PbSe, PbS, etc. and at the bottom of conduction band of $Be_2Te_3$, $Si_{1-x}Ge_x$, etc., and cases for linear bands that happens at the $K$ point of graphene, the $L$ point of certain $Bi_{1-x}Sb_x$, are not discussed. All these previous achievement are suitable for study the electronic transport of a certain material, but not suitable for the materials-searching and the conditions-optimization problems of TE/electronics.

In this paper, we have derived the analytical form of anisotropic transport distribution tensor for a parabolic band in a three-dimensional materials system under the condition that the relaxation time tensor is only a function of energy at a certain temperature, i.e. $\boldsymbol{\tau}(\mathbf{k}) = \tau_0(E)\mathbf{T}$, where $\mathbf{T}$ is a constant tensor. We have found that the transport distribution tensor can be asymmetrical, i.e. exhibiting a deviation from the Onsager relation, which is usually not considered in previous literatures on modeling electronic transport. We have then developed the analytical form of anisotropic transport distribution tensor for a linear band, as well as a non-parabolic band of a more general form beyond the Lax model $E + E^2/E_g = (\hbar^2/2)\mathbf{k}^T \mathbf{M}^{-1} \mathbf{k}$ [7], where $E_g$ is the direct band gap. Then, we have generalized the results to two- and one-dimensional materials systems. Furthermore, we have derived the anisotropic distribution tensor under the condition that the relaxation time tensor is a function of not only energy, but also carrier velocity. Finally, we have done a comparison between the numerical method used in previous literatures and our analytical results reported in this paper.

For clarification of the problem, we will discuss the transport distribution tensor in the Cartesian coordinates system that coincide with the principal axes of the effective-mass tensor, e.g. for a parabolic band $E(\mathbf{k}) = (\hbar^2/2)\mathbf{k}^T \mathbf{M}^{-1}\mathbf{k}$, where $\mathbf{k}$ is the lattice momentum measured from the bottom/top of the conduction/valence band, and $\mathbf{M}$ has no off-diagonal components. $\Xi(E)$ expressed in other coordinates systems can be easily obtained by rotation matrixes. For an isotropic parabolic band, the dispersion reduces to $E(\mathbf{k}) = (\hbar^2/2m)\mathbf{k}^2$, where $\mathbf{k}^2 = k_1^2 + k_2^2 + k_3^2$. By symmetry, we know that $\Xi(E)$ can only take the form of $\Xi(E) = \Xi_0(E)\mathscr{U}$, where $\Xi_0(E)$ is a scalar function of $E$, and $\mathscr{U}$ is the unitary matrix. Hence, we know

$$\Xi(E) = \sum_{\mathbf{k}} \delta(E - E_{\mathbf{k}}) \tau_0(E) \begin{pmatrix} v_1^2 & & \\ & v_2^2 & \\ & & v_3^2 \end{pmatrix}. \tag{7}$$

By symmetry, we have

$$\Xi_0(E) = \Xi_{ii}(E) = \frac{1}{3}\sum_{\mathbf{k}} \delta(E - E_{\mathbf{k}}) \tau_0(E) v^2 = \frac{2E}{3m} D(E) \tau_0(E), \tag{8}$$

where $v^2 = v_1^2 + v_2^2 + v_3^2$, $i=1,2$ or $3$, and $D(E) = \sum_{\mathbf{k}} \delta(E - E_{\mathbf{k}})$ is the density of states, which is $(1/4\pi^2)(2m/\hbar^2)^{3/2} E^{1/2}$ for a 3D isotropic parabolic band. Thus, we have

$$\Xi_0(E) = (2^{1/2}/3\pi^2\hbar^3) m^{1/2} E^{3/2} \tau_0(E). \tag{9}$$

For a general anisotropic parabolic dispersion expressed in the principal-axes coordinates, $E(\mathbf{k}) = (\hbar^2/2)(k_1^2/m_1 + k_2^2/m_2 + k_3^2/m_3)$, the transport distribution tensor is $\Xi(E) = \sum_{\mathbf{k}} \delta(E - E_{\mathbf{k}}) \mathbf{v} \otimes \boldsymbol{\tau}\mathbf{v}$, which is generally not equal to $\sum_{\mathbf{k}} \delta(E - E_{\mathbf{k}}) \boldsymbol{\tau}\mathbf{v} \otimes \mathbf{v}$, i.e. $v_i \sum_l \tau_{jl} v_l \neq v_j \sum_l \tau_{il} v_l$ for $i \neq j$ in an anisotropic case. Thus, to get the transport distribution tensor, we need to calculate each $\sum_{\mathbf{k}} \delta(E - E_{\mathbf{k}}) v_i v_j \tau_{pq}$, which we define as $\Theta_{ijpq}(E)$, where we have $\Xi_{ij}(E) = \sum_l \Theta_{iljl}(E)$. $\Theta_{ijpq}(E)$ can then be treated as the mean value of $v_i v_j \tau_{pq}$ at the constant energy surface of $E$, multiplied by the density of states at $E$, i.e. $\Theta_{ijpq}(E) = \tau_{pq} \langle v_i v_j \rangle_E D(E)$. According to $v_i = \hbar^{-1} \partial E/\partial k_i = \hbar k_i/m_i$, we have $\langle v_i v_j \rangle_E = (\hbar^2/m_i m_j) \langle k_i k_j \rangle_E = \delta_{ij} 2E/3m_i$. Thus, $\Theta_{ijpq}(E) = \delta_{ij} 2\tau_{pq} D(E) E/3m_i$ and $\Xi_{ij}(E) = (2E/3m_i) D(E) \tau_{ji}$.

Hence, we have found that if the relaxation time tensor $\tau$ does not diagonalize in the same coordinates as the dispersion relation $E(\mathbf{k})$ does, the off-diagonal components of the transport distribution tensor $\Xi(E)$ will not be cancelled out, i.e. $\Xi_{ij}(E) = (2E/3m_i)D(E)\tau_{ji}$, which is an advanced explanation for the assumptions used in Ref. . We also see that even if the relaxation time tensor $\tau$ is a symmetric tensor, the transport distribution tensor $\Xi$ is not necessarily symmetric, if $\tau$ does not diagonalizes in the same coordinates system as $E(\mathbf{k})$, i.e. $\Xi_{ij}(E) = (2E/3m_i)D(E)\tau_{ji}$ is generally not equal to $\Xi_{ji}(E) = (2E/3m_j)D(E)\tau_{ij}$ ($i \neq j$). This possible deviation from the Onsager relation has not been considered in previous literatures on anisotropic electronic transport, which simply assumed that if $\tau$ is symmetrical, $\Xi$ is naturally symmetrical. Our finding of possible deviation from Onsager's relation here is consistent with the work done by Truesdell [8] and Bies [6] through different approaches.

For low-dimensional system, the transport distribution tensor can be derived in the similar way. For a 2D isotropic parabolic band, the transport distribution can be obtained by the same symmetry argument we did above for 3D case, except that $\sum_{\mathbf{k}_{[2D]}} \delta(E - E_{\mathbf{k}_{[2D]}})\tau_{0,[2D]}(E)v_{[2D]}^2$ should be divided by 2 instead of 3. Thus, we have

$$\Xi_{0,[2D]}(E) = \Xi_{[2D],ii}(E) = \frac{1}{2}\sum_{\mathbf{k}_{[2D]}} \delta(E - E_{\mathbf{k}_{[2D]}})\tau_{0,[2D]}(E)v_{[2D]}^2 = (E/m)D_{[2D]}(E)\tau_{0,[2D]}(E). \tag{10}$$

For a 2D anisotropic parabolic band, we have $\Xi_{[2D],ij}(E) = \sum_l \Theta_{[2D],iljl}(E)$, where

$$\Theta_{[2D],ijpq}(E) = \sum_{\mathbf{k}_{[2D]}} \delta(E - E_{\mathbf{k}_{[2D]}})v_{[2D],i}v_{[2D],j}\tau_{[2D],pq}. \tag{11}$$

$\Theta_{[2D],ijpq}(E)$ can also then be treated as the mean value of $v_{[2D],i}v_{[2D],j}\tau_{[2D],pq}$ at the constant energy circle of $E$, multiplied by the density of states at $E$, in the 2D system, i.e. $\Theta_{[2D],ijpq}(E) = \tau_{[2D],pq}\langle v_{[2D],i}v_{[2D],j}\rangle_E D_{[2D]}(E)$. Then we have

$$\langle v_{[2D],i}v_{[2D],j}\rangle_E = (\hbar^2/m_im_j)\langle k_{[2D],i}k_{[2D],j}\rangle_E = \delta_{ij}E/m_i. \tag{12}$$

Thus, $\Theta_{[2D],ijpq}(E) = \delta_{ij}\tau_{[2D],pq}D_{[2D]}(E)E/m_i$ and $\Xi_{[2D],ij}(E) = (E/m_i)D_{[2D]}(E)\tau_{[2D],ji}$. For the 1D parabolic band, we simply have,

$$\Xi_{[1D]}(E) = (2E/m)D_{[1D]}(E)\tau_{0,[1D]}(E). \tag{13}$$

To describe a non-parabolic band, the Lax model, i.e. $E + E^2/E_g = (\hbar^2/2)\mathbf{k}^T\mathbf{M}^{-1}\mathbf{k}$, is often used, e.g. the *L*-point band edges of bismuth and bismuth antimony. Instead of talking about this specific form of non-parabolic band, we discuss a general form of non-parabolic band defined as,

$$\tilde{E} = \sum_{N=0}^{\infty} c_N E^N = (\hbar^2/2)\mathbf{k}^T\mathbf{M}^{-1}\mathbf{k}, \tag{14}$$

where $c_N$ are constants. In the principal-axes coordinates of $\tilde{E}(\mathbf{k})$, we have

$$\tilde{E} = \sum_{N=0}^{\infty} c_N E^N = (\hbar^2/2)(k_1^2/m_1 + k_2^2/m_2 + k_3^2/m_3). \tag{15}$$

The transport distribution tensor can still be calculated as $\Xi_{ij}(E) = \sum_l \tau_{jl} \langle v_i v_l \rangle_E D(E)$, except that

$$\langle v_i v_l \rangle_E D(E) = \langle v_i v_l \rangle_{\tilde{E}(E)} D(\tilde{E}) \frac{d\tilde{E}}{dE} \tag{16}$$

and

$$v_i = \frac{1}{\hbar}\frac{\partial E}{\partial k_i} = \frac{1}{\hbar}\frac{\partial \tilde{E}}{\partial k_i}\Big/\frac{d\tilde{E}}{dE} = \frac{\hbar k_i}{m_i}\Big/\frac{d\tilde{E}}{dE}. \tag{17}$$

Thus, we have

$$\langle v_i v_j \rangle_E D(E) = \delta_{ij} \frac{2\tilde{E}}{3m_i} D(\tilde{E})\Big/\frac{d\tilde{E}}{dE} \tag{18}$$

and

$$\Xi_{ij}(E) = \sum_l \tau_{jl} \delta_{il} \frac{\hbar^2}{m_i m_l} D(\tilde{E})\Big/\frac{d\tilde{E}}{dE} = \tau_{ji} \frac{2\tilde{E}}{3m_i} D(\tilde{E})\Big/\frac{d\tilde{E}}{dE}. \tag{19}$$

For a 2D non-parabolic band, where $\tilde{E} = \sum_{N=0}^{\infty} c_{[2D],N} E^N = (\hbar^2/2)(k_{[2D],1}^2/m_1 + k_{[2D],2}^2/m_2)$, we have

$$\langle v_{[2D],i} v_{[2D],j} \rangle_E D_{[2D]}(E) = \delta_{ij} \frac{\tilde{E}}{m_i} D_{[2D]}(\tilde{E})\Big/\frac{d\tilde{E}}{dE} \tag{20}$$

and

$$\Xi_{[2D],ij}(E) = \tau_{[2D],ji} \frac{\tilde{E}}{m_i} D_{[2D]}(\tilde{E})\Big/\frac{d\tilde{E}}{dE}. \tag{21}$$

For a 1D non-parabolic band, where $\tilde{E} = \sum_{N=0}^{\infty} c_{[1D],N} E^N = (\hbar^2/2)(k_{[1D]}^2/m)$, we have

$$\Xi_{[1D]}(E) = \frac{2\tilde{E}}{m_i} D_{[1D]}(\tilde{E}) \tau_{[1D]}(E)\Big/\frac{d\tilde{E}}{dE}. \tag{22}$$

The linear band case, e.g. the possible Dirac point in bismuth antimony, is included in our defined general form of non-parabolic dispersion, if we take $c_N = c_2 \delta_{2,N}$, i.e.

$$E = (\hbar / \sqrt{2c_2}) \sqrt{k_x^2/m_1 + k_y^2/m_2 + k_z^2/m_3} \tag{23}$$

and

$$\Xi_{ij}(E) = \frac{1}{6\pi^2 \hbar^3} \frac{v_i^2}{v_1 v_2 v_3} E^2 \tau_{ji}, \tag{24}$$

where $v_i = 1/\sqrt{2c_2 m_i}$ and $m_i$ is a parameter defined in Eq. (15), which does not mean mass anymore in the linear dispersion.

.

For 2D Dirac cones, we have

$$E = (\hbar / \sqrt{2c_{[2D],2}}) \sqrt{k_1^2/m_1 + k_2^2/m_2} \tag{25}$$

and

$$\Xi_{[2D],ij}(E) = \frac{v_{[2D],i}^2}{4\pi \hbar^2 v_{[2D],1} v_{[2D],2}} \tau_{[2D],ji} E, \tag{26}$$

where $v_{[2D],i} = 1/\sqrt{2c_{[2D],2} m_i}$. The dispersion relation reduces to $E = \hbar k / \sqrt{2mc_{[2D],2}}$ in isotropic Dirac cones, e. g. in graphene and topological insulators, where the carrier group velocity is $v = 1/\sqrt{2mc_{[2D],2}}$ and the transport distribution tensor is $\Xi_{[2D]}(E) = (E/4\pi\hbar^2)\tau_{[2D]}$. Following the similar argument, we can obtain the transport distribution for a 1D system as

$$\Xi_{[1D]}(E) = v_{[1D]} \tau_{[1D]} / \pi \hbar. \tag{27}$$

All the above discussions are valid for cases where $\tau$ is only a function of energy E, and not a function of velocity $\mathbf{v}$. For some systems where semi-empirical constant mean free path approximation is preferred, especially in low-dimensional systems [9-12]. Thus, we will start from the low-dimensional systems. For an 1D system, the transport distribution is

$$\Xi_{[1D]}(E) = \sum_{\mathbf{k}_{[1D]}} \delta(E - E_{\mathbf{k}_{[1D]}}) \mathbf{v}_{[1D]} \otimes \lambda_{[1D]}(E)(\mathbf{v}_{[1D]}/|\mathbf{v}_{[1D]}|), \tag{28}$$

where we assumed that the mean free path $\lambda_{[1D]}$ is only a function of $E$ at a specific temperature. Everything reduces to scalar in a 1D system, i.e.

$$\Xi(E) = \sum_k \delta(E - E_k) |v| \lambda(E) = \lambda(E) \sqrt{2E/m} D_{[1D]}(E). \tag{29}$$

For a 2D system, the 2D transport distribution tensor is

$$\Xi_{[2D]}(E) = \sum_{\mathbf{k}_{[2D]}} \delta(E - E_{\mathbf{k}_{[2D]}})\mathbf{v}_{[2D]} \otimes \lambda_{[2D]}(\mathbf{v}_{[2D]}/|\mathbf{v}_{[2D]}|). \tag{30}$$

Thus, we want to calculate each $\sum_{\mathbf{k}_{[2D]}} \delta(E - E_{\mathbf{k}_{[2D]}})v_{[2D],i}v_{[2D],j}(\lambda_{[2D],pq}/\sqrt{v_{[2D],1}^2 + v_{[2D],2}^2})$

By calculation, we have found that

$$\sum_{\mathbf{k}_{[2D]}} \delta(E - E_{\mathbf{k}_{[2D]}})v_{[2D],i}v_{[2D],i} \frac{\lambda_{[2D],pq}}{\sqrt{v_{[2D],1}^2 + v_{[2D],2}^2}} = \lambda_{[2D],pq} \frac{4\sqrt{2E}}{\sqrt{m_i}} \frac{m_i \cdot EK(\sqrt{1-\frac{m_i}{m_j}}) - m_j \cdot EE(\sqrt{1-\frac{m_i}{m_j}})}{m_i - m_j} D_{[2D]}(E) \tag{31}$$

and

$$\sum_{\mathbf{k}_{[2D]}} \delta(E - E_{\mathbf{k}_{[2D]}})v_{[2D],i}v_{[2D],j} \frac{\lambda_{[2D],pq}}{\sqrt{v_{[2D],1}^2 + v_{[2D],2}^2}} = 0 \ (i \neq j), \tag{32}$$

where *EK* and *EE* are the first type and second type elliptical integrations. Thus, we have

$$\Xi_{[2D],ij}(E) = \lambda_{[2D],ji} \frac{4\sqrt{2E}}{\sqrt{m_i}} \frac{m_i \cdot EK(\sqrt{1-\frac{m_i}{m_j}}) - m_j \cdot EE(\sqrt{1-\frac{m_i}{m_j}})}{m_i - m_j} D_{[2D]}(E). \tag{33}$$

For 3D system, we have $\Xi(E) = \sum_{\mathbf{k}} \delta(E - E_{\mathbf{k}})\mathbf{v} \otimes \lambda(\mathbf{v}/|\mathbf{v}|)$, this is used for systems such as $Bi_2Te_3$ and $Sb_2Te_3$ [3]. We need to calculate,

$$\sum_{\mathbf{k}} \delta(E - E_{\mathbf{k}})v_i v_j \frac{\lambda_{pq}}{\sqrt{v_1^2 + v_2^2 + v_3^2}} = \lambda_{pq} \left\langle \frac{v_i v_j}{\sqrt{v_1^2 + v_2^2 + v_3^2}} \right\rangle_E D(E). \tag{34}$$

By calculations, we have,

$$\left\langle \frac{v_i v_j}{\sqrt{v_1^2 + v_2^2 + v_3^2}} \right\rangle_E = \delta_{ij} \frac{\sqrt{2E}}{m_i} \sqrt{\tilde{m}}, \tag{35}$$

where we defined

$$\sqrt{\tilde{m}} = \left\langle \frac{\sin^2\theta\cos^2\varphi}{\sqrt{\frac{\sin^2\theta\cos^2\varphi}{m_1} + \frac{\sin^2\theta\sin^2\varphi}{m_2} + \frac{\cos^2\theta}{m_3}}} \right\rangle_E = \int_{\varphi=0}^{2\pi} \int_{\theta=0}^{\pi} \frac{\sin^2\theta\cos^2\varphi}{\sqrt{\frac{\sin^2\theta\cos^2\varphi}{m_1} + \frac{\sin^2\theta\sin^2\varphi}{m_2} + \frac{\cos^2\theta}{m_3}}} \sin\theta d\theta d\varphi , \tag{36}$$

which turns out to have non-elementary functions in the analytical form. Putting Eq. (35) into Eq. (34) we have

$$\lambda_{pq}\sum_{\mathbf{k}}\delta(E-E_{\mathbf{k}})\frac{v_i v_j}{\sqrt{v_1^2+v_2^2+v_3^2}} = \delta_{ij}\lambda_{pq}\frac{\sqrt{2E}}{m_i}\sqrt{\tilde{m}}D(E), \tag{37}$$

which gives

$$\Xi_{ij}(E) = \lambda_{ji}\frac{\sqrt{2E}}{m_i}\sqrt{\tilde{m}}D(E). \tag{38}$$

Up to now we have considered most of the situations for anisotropic transport distribution tensor, including parabolic, non-parabolic and linear dispersions, including three-, two- and one-dimensional systems, and also including relaxation time approximation and mean free path approximation in the present work. We now compare the numerical method used in previous literature and the analytical method we have developed here in the present paper in some specific materials. The numerical method basically uses a normalized quasi-delta function $\sigma^{-1}\tilde{\delta}(E/\sigma)$ to mimic the delta function in Eq. (6) [13-15]. $\sigma^{-1}\tilde{\delta}(E/\sigma)$ is a smeared convolution of $\delta(E)$ and $\sigma$ is the smearing, i.e. $\int_{-\infty}^{\infty}\sigma^{-1}\tilde{\delta}(E/\sigma)dE = 1$ and $\lim_{\sigma\to 0}\sigma^{-1}\tilde{\delta}(E/\sigma) = \delta(E)$. Usually, the Gaussian smearing function $\sigma^{-1}\tilde{\delta}(E/\sigma) = (1/\sqrt{2\pi}\sigma)\exp(-E^2/2\sigma^2)$ is chosen for its simplicity and special integrational properties, and the smearing is set to be equal to the thermal smearing of $\sigma = k_B T$. A grid of points in the *k*-space are sampled. Thus, the density of states and the transport distribution tensor can be approximated as [13-15],

$$D(E) = \sum_{\mathbf{k}\in\{\text{Sampled Grid}\}}\sigma^{-1}\tilde{\delta}[(E-E_{\mathbf{k}})/\sigma] \tag{39}$$

and

$$\Xi(E) = \sum_{\mathbf{k}\in\{\text{Sampled Grid}\}}\sigma^{-1}\tilde{\delta}[(E-E_{\mathbf{k}})/\sigma]\mathbf{v}\otimes\boldsymbol{\tau}\mathbf{v} \tag{40}$$

We consider the situation under relaxation time approximation. For the case of a 3D anisotropic parabolic band valley, we illustrate the *T*-point valence band valley of bulk bismuth [16], as shown in Fig. 1. We see that for both density of states and the components of transport distribution tensor, the results from our analytical method are consistent with the results from numerical methods used by previous literatures. Furthermore, the numerical method become less trustable when it is close to a band-edge, because the density of states and the transport distribution are broadened by the smeared quasi-delta function, which can barely capture the sudden change of density of states or transport distribution at the band edge. However, the discontinuity of density of states and transport distribution at band edges is very essential for thermoelectrics and electronics [17]. Thus, we propose that we should use the analytical method as much as possible to increase the accuracy of thermoelectric modeling. For the case of a 3D anisotropic non-parabolic band valley and linear band valley, we illustrate the *L*-point electron valley of bulk PbTe [18] and the Dirac point of bulk $Bi_{0.96}Sb_{0.04}$ [19], respectively, as shown in Fig. 2 (a) and (b).

For 2D materials, we illustrate the 2D parabolic conduction band valley at the *K* point of $MoS_2$ [20], as shown in Fig. 3 (a). Researchers have studied anisotropic Dirac cones in graphene superlattice and in $Bi_{1-x}Sb_x$ thin films. We illustrate the anisotropic Dirac cone in graphene supperlattice studied in Ref. [21, 22], and the anisotropic Dirac cone in bisectrix oriented $Bi_{0.96}Sb_{0.04}$ thin film studied in Ref. [19, 23], as shown in Fig. 3 (b) and (c), respectively.

Lastly, we illustrate the single valley in carbon nanoribon for 1D systems. For a semiconducting carbon nanoribbon, the band edge is parabolic with an effective mass. We illustrate the metallic armchar carbon nanoribbon with a width of 6.02 nm and Dirac fermion group velocity of $8 \times 10^5$ m/s [24, 25], as shown in Fig. 4 (a), where the 1D density of states and the 1D transport distribution function remains a constant when the carrier energy is greater than zero. For a semiconducting carbon nanoribbon, there might form a 1D linear dispersion relation at the band edge, e.g. a metallic armchair nanoribbon with a width of 21 nm, a band gap of 0.65 eV and an effective mass of $0.05m_e$ [24, 26], as show in Fig. 4(b). We noticed that for the semiconducting carbon nanoribbon, though the density of states diverges at the band edge, the transport distribution converges to 0, which is why the electrical conductivity is still finite.

In conclusion, we have derived the analytical forms of anisotropic transport distribution tensor for parabolic, non-parabolic, and linear valleys, in 3D, 2D and 1D materials systems, under both the relaxation time approximation and the mean free path approximation. We have found that the Onsager relation for electronic transport can be deviated, if the relaxation time tensor does not diagonalize in the same coordinates frame which diagonalizes the effective mass tensor. We then calculated the transport distribution function for some band valleys of several interesting materials systems, including the T-point hole valley of bulk bismuth, the L-point electron valley of bulk PbTe, the anisotropic Dirac point in bulk $Bi_{0.96}Sb_{0.04}$, the *K* point electron valley of $MoS_2$, the anisotropic Dirac cones in graphene superlattice and $Bi_{1-x}Sb_x$ thin films, and also the single valley in semiconducting and metallic carbon nanoribbons.

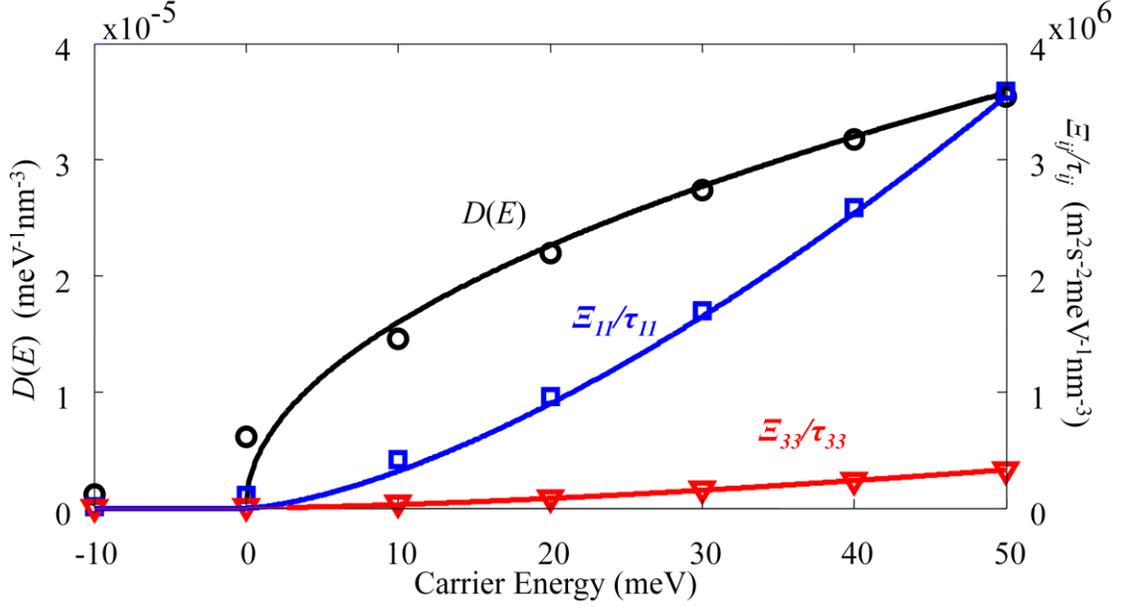

Fig. 1: Comparison between the density of states (black), principal components of transport distribution tensor (blue and red), calculated from the analytical method developed in this paper (solid lines) and the numerical method used in previous literatures (dots). The valence band valley at the $T$ point of bulk bismuth is illustrated as an example. The principal effective masses used for calculations are $m_1=m_2=0.059m_e$ and $m_3=0.634m_e$ [16], where $m_e$ is the free electron mass. The components of transport distribution tensor are normalized by corresponding components of the relaxation time tensor for generality. In the numerical integration, the Gaussian form of $\sigma^{-1}\tilde{\delta}(E/\sigma)=(1/\sqrt{2\pi}\sigma)\exp(-E^2/2\sigma^2)$ [13-15] is used to be the quasi-delta function, and the smearing is set to be $k_B T$, where $T=100$ K. The sampled $\boldsymbol{k}$-space grid is evenly distributed in the $\boldsymbol{k}$-space, and set to have a density of $(2.5\times 10^6)^3/m^{-3}$.

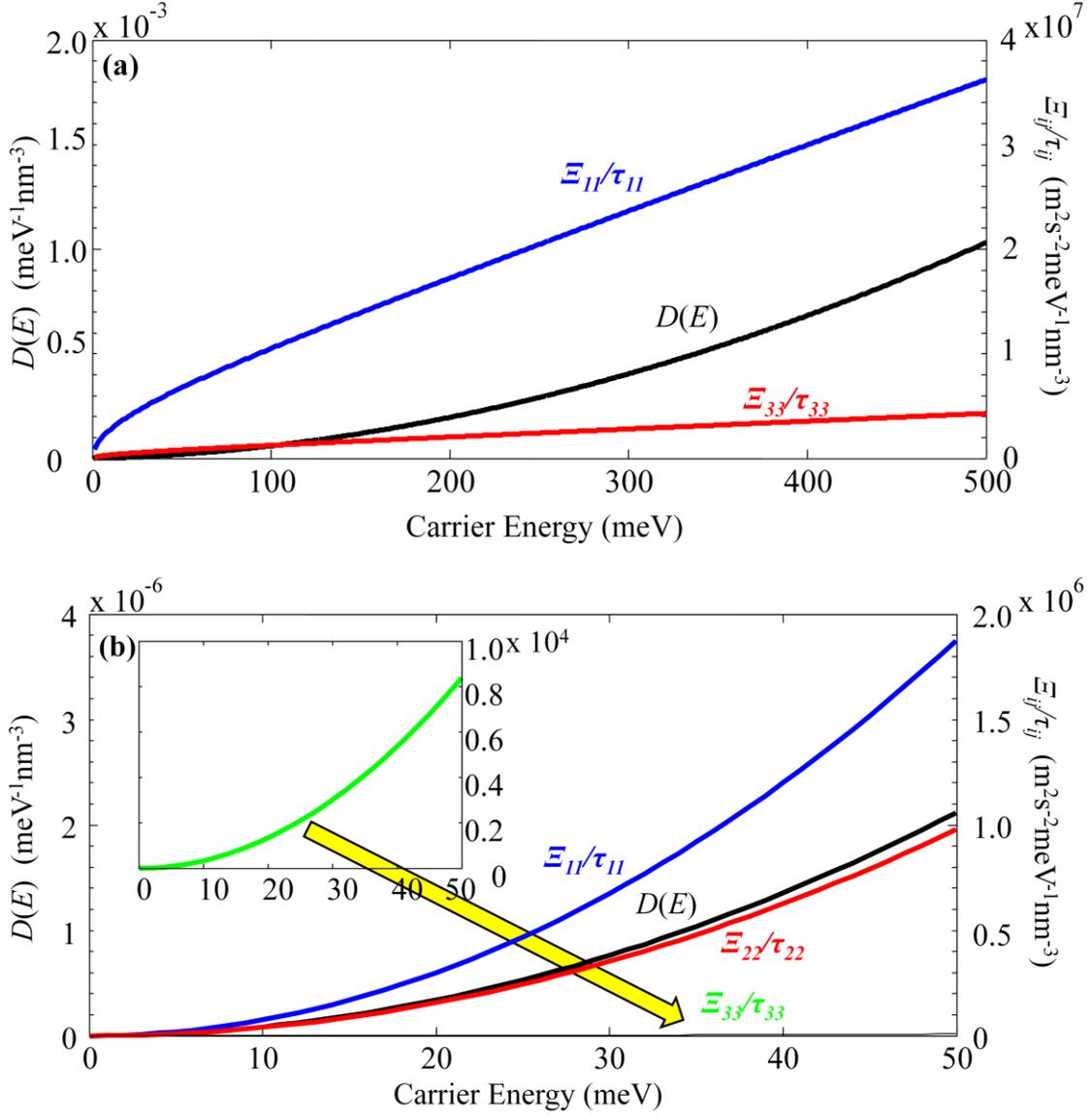

Fig. 2: Density of states and principal components of transport distribution tensor of (a) the non-parabolic conduction band valley at the $L$ point of bulk PbTe, and (b) the Dirac point in bulk $Bi_{0.96}Sb_{0.04}$. (a) The principal effective masses used for calculations are $m_1=m_2=0.06m_e$ and $m_3=0.505m_e$ [18], and the non-parabolic form $E+E^2/E_g = \tilde{E}$ is used, where $E_g$=189.7 meV [18]. (b) The principal group velocities used for calculations are $v_1= 1.63\times10^6$ m/s, $v_2= 1.18\times10^6$ m/s and $v_3= 1.09\times10^5$ m/s [19].

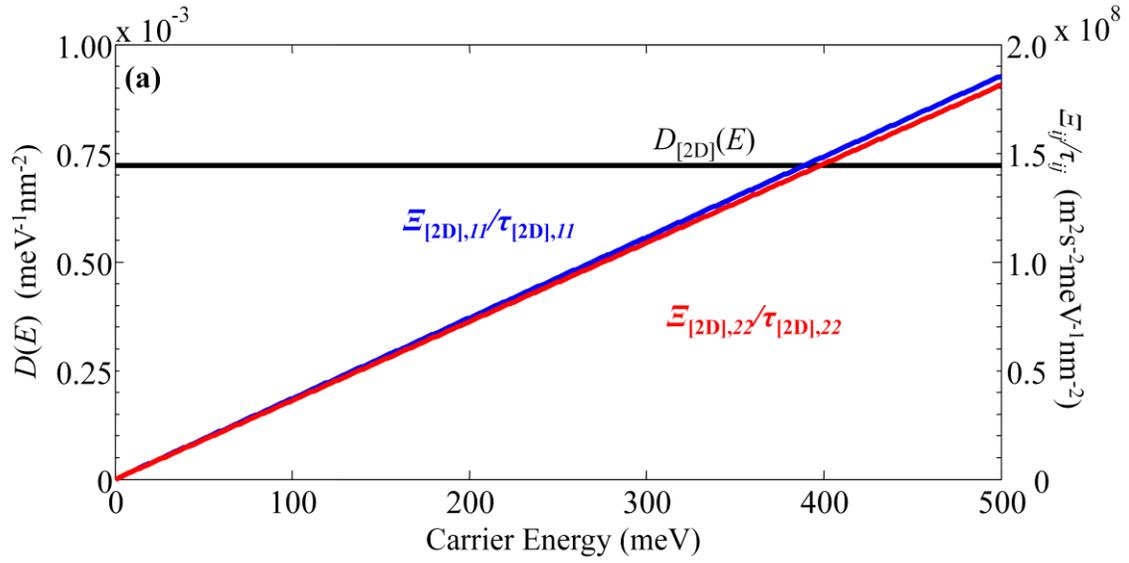

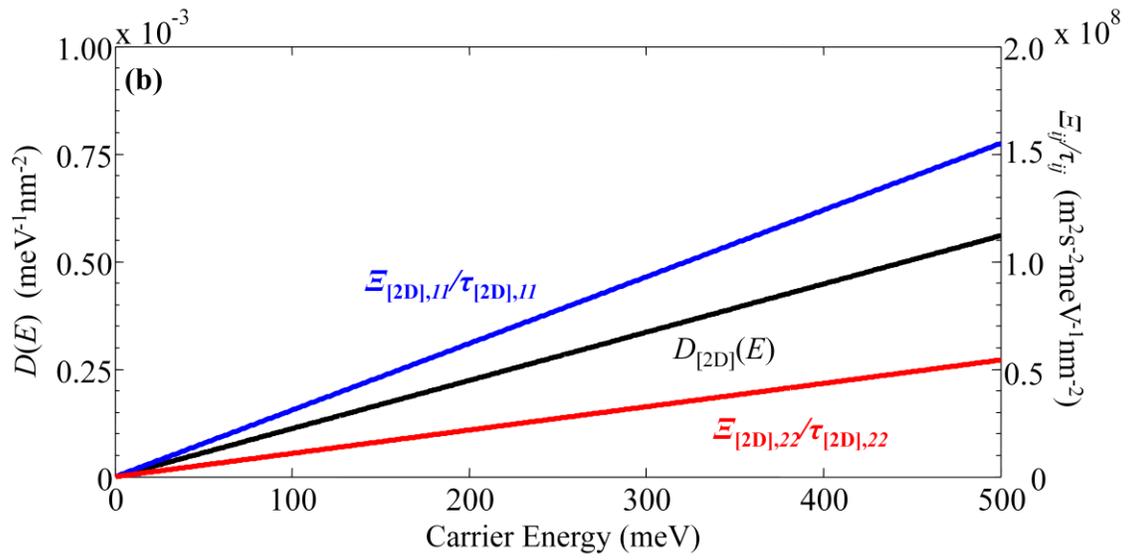

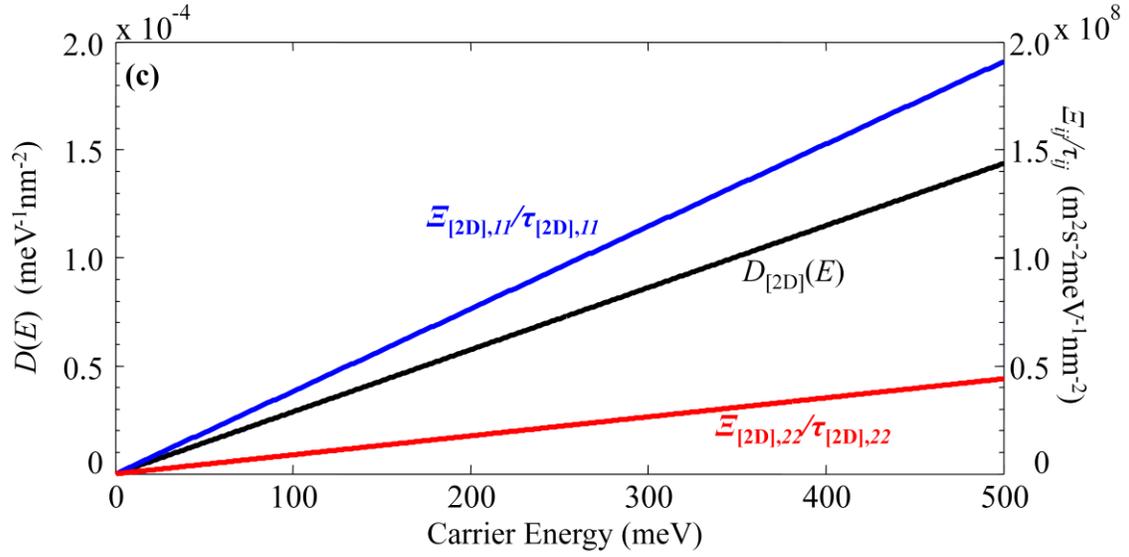

Fig. 3: Density of states and principal components of transport distribution tensor for two-dimensional systems. (a) The parabolic conduction band valley at the $K$ point of $MoS_2$ [20]. (b) The anisotropic Dirac cone in graphene supperlattice [21, 22]. (c) The anisotropic Dirac cone in bisectrix oriented $Bi_{0.96}Sb_{0.04}$ thin films [19, 23].

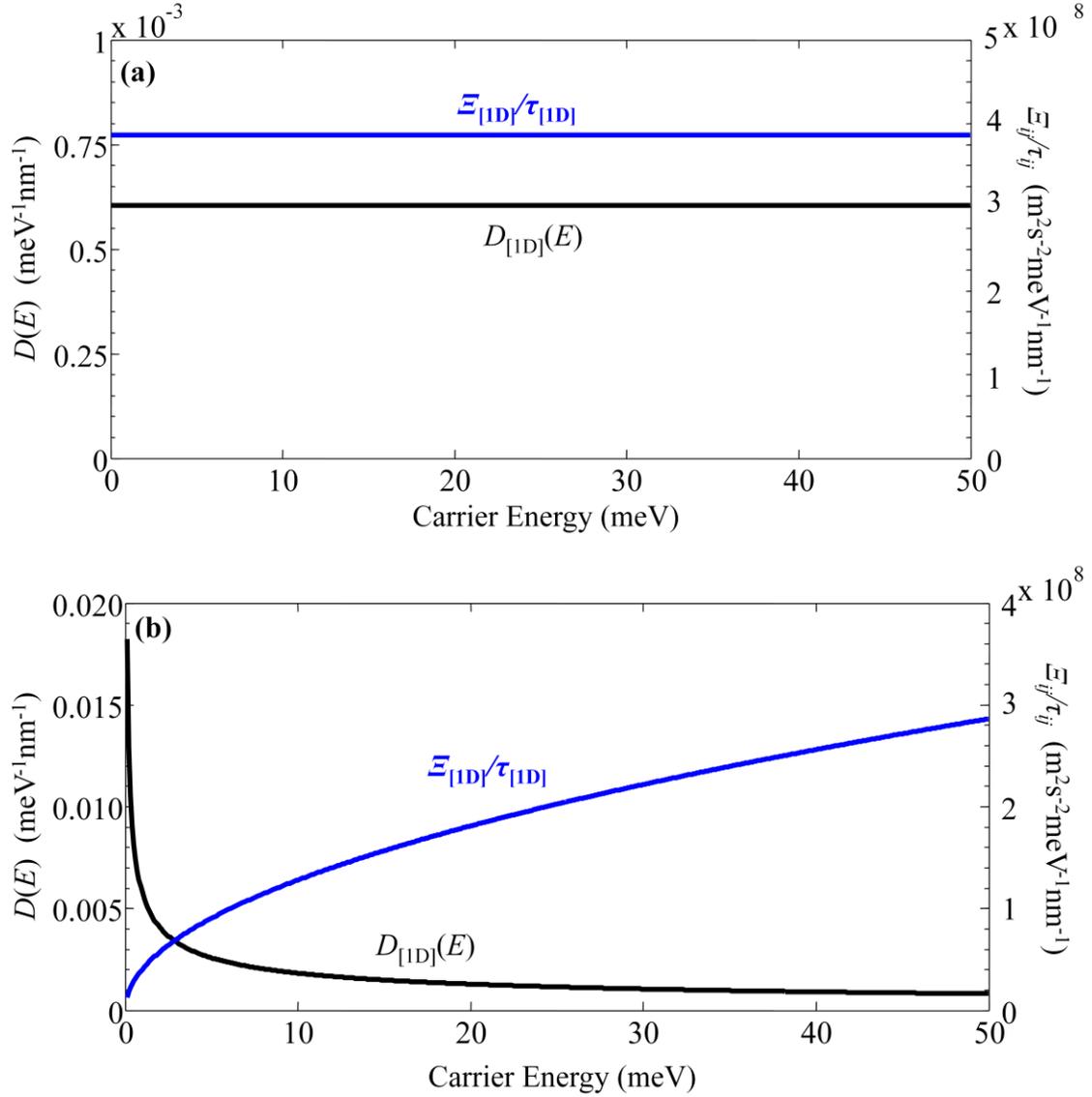

Fig. 4: Density of states and principal components of transport distribution tensor for one-dimensional systems. (a) Metallic armchair carbon nanoribbon with a width of 6.02 nm and Dirac fermion group velocity of $8\times10^5$ m/s [24, 25]. (b) Semiconducting armchair nanoribbon with a width of 21 nm, a band gap of 0.65 eV and an effective mass of $0.05m_e$ [24, 26].

1. T. Scheidemantel, C. Ambrosch-Draxl, T. Thonhauser, J. Badding, and J. Sofo, Phys. Rev. B **68** (2003).
2. S. Lee and P. von Allmen, Appl. Phys. Lett. **88**, 022107 (2006).
3. B. Y. Yavorsky, N. Hinsche, I. Mertig, and P. Zahn, Phys. Rev. B **84**, 165208 (2011).
4. T. Teramoto, T. Komine, M. Kuraishi, R. Sugita, Y. Hasegawa, and H. Nakamura, J. Appl. Phys. **103**, 043717 (2008).
5. T. Teramoto, T. Komine, S. Yamamoto, M. Kuraishi, R. Sugita, Y. Hasegawa, and H. Nakamura, J. Appl. Phys. **104**, 053714 (2008).
6. W. Bies, R. Radtke, H. Ehrenreich, and E. Runge, Phys. Rev. B **65** (2002).
7. B. Lax and J. G. Mavroides eds., (Academic Press, New York, 1960).
8. C. Truesdell, *Rational Thermoelectrics: A Course of Lectures on Selected Topics* (McGraw-Hill, New York, 1969).
9. V. Sandomirskii, Soviet Journal of Experimental and Theoretical Physics **25**, 101 (1967).
10. L. Hicks and M. Dresselhaus, Phys. Rev. B **47**, 12727 (1993).
11. L. Hicks and M. Dresselhaus, Phys. Rev. B **47**, 16631 (1993).
12. E. Rogacheva, S. Grigorov, O. Nashchekina, S. Lyubchenko, and M. Dresselhaus, Appl. Phys. Lett. **82**, 2628 (2003).
13. C.-L. Fu and K.-M. Ho, Phys. Rev. B **28**, 5480 (1983).
14. R. J. Needs, R. M. Martin, and O. H. Nielsen, Phys. Rev. B **33**, 3778 (1986).
15. N. Mazari, PhD Thesis, University of Camridge, 1996.
16. R. J. Dinger and A. W. Lawson, Phys. Rev. B **7**, 5215 (1973).
17. G. Mahan and J. Sofo, Proceedings of the National Academy of Sciences **93**, 7436 (1996).
18. S. Yuan, G. Springholz, G. Bauer, and M. Kriechbaum, Phys. Rev. B **49**, 5476 (1994).
19. S. Tang and M. S. Dresselhaus, Nano Lett. **12**, 2021 (2012).
20. T. Cheiwchanchamnangij and W. R. Lambrecht, Phys. Rev. B **85**, 205302 (2012).
21. C.-H. Park, L. Yang, Y.-W. Son, M. L. Cohen, and S. G. Louie, Nature Physics **4**, 213 (2008).
22. S. Rusponi, et al., Phys. Rev. Lett. **105**, 246803 (2010).
23. S. Tang and M. S. Dresselhaus, Nanoscale **4**, 7786 (2012).
24. H. Raza and E. C. Kan, Phys. Rev. B **77**, 245434 (2008).
25. A. Naeemi and J. D. Meindl, in *Interconnect Technology Conference, 2008. IITC 2008. International* (IEEE, 2008), p. 183.
26. G. Liang, N. Neophytou, D. E. Nikonov, and M. S. Lundstrom, Electron Devices, IEEE Transactions on **54**, 677 (2007).